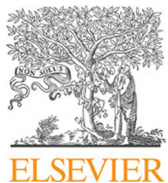
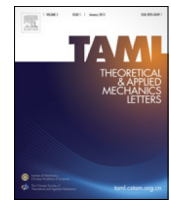

Letter

# Two-dimensional modeling of the self-limiting oxidation in silicon and tungsten nanowires

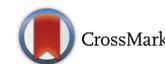

Mingchao Liu [a,b], Peng Jin [a], Zhiping Xu [a], Dorian A.H. Hanaor [b], Yixiang Gan [b], Changqing Chen [a,*]

[a] *Department of Engineering Mechanics, CNMM & AML, Tsinghua University, Beijing 100084, China*
[b] *School of Civil Engineering, The University of Sydney, Sydney, NSW 2006, Australia*

## HIGHLIGHTS

- A new diffusion-controlled kinetic model for nanowire oxidation is developed.
- A finite reactive region is included to account for oxidation stress and suboxide formation.
- Self-limiting nanowire oxidation and its curvature/temperature dependence are predicted.
- Results are consistent with observed oxidation behavior of silicon (Si) and tungsten (W) nanowires.

ABSTRACT

Self-limiting oxidation of nanowires has been previously described as a reaction- or diffusion-controlled process. In this letter, the concept of finite reactive region is introduced into a diffusion-controlled model, based upon which a two-dimensional cylindrical kinetics model is developed for the oxidation of silicon nanowires and is extended for tungsten. In the model, diffusivity is affected by the expansive oxidation reaction induced stress. The dependency of the oxidation upon curvature and temperature is modeled. Good agreement between the model predictions and available experimental data is obtained. The developed model serves to quantify the oxidation in two-dimensional nanostructures and is expected to facilitate their fabrication via thermal oxidation techniques.

© 2016 The Author(s). Published by Elsevier Ltd on behalf of The Chinese Society of Theoretical and Applied Mechanics. This is an open access article under the CC BY-NC-ND license (http://creativecommons.org/licenses/by-nc-nd/4.0/).

Owing to their unique physical and chemical properties, silicon nanowires (Si NWs) are a promising candidate for a wide range of applications [1–5]. Thermal oxidation is one of the most fundamental processes used for manufacturing nanostructured devices in general and Si NWs in particular, allowing accurate control of the size, surface structure, and electronic properties [6–9]. Existing experimental studies have shown that the oxidation behavior of silicon nanostructures differs significantly from that of bulk Si [10]. In particular, the oxidation kinetics of Si NWs are found to be surface curvature- and temperature-dependent [11]. The initial oxidation rate of Si NWs with small diameters can be significantly higher when compared to the planar oxide growth on Si. Depending on their initial size, the oxidation of Si NWs may become immeasurably slow during prolonged oxidation studies. This phenomena is referred to as retarded oxidation [12,13], or self-limiting oxidation [14–17], and is more pronounced at lower temperatures [8,11]. However, understanding the self-limiting oxidation in nanowires is hindered by the complex coupling between mechanical deformation and oxidation, making it difficult to apply the self-limiting effect to control the size of Si NWs in fabrication processes [8,18,19]. To facilitate improved manufacturing quality of Si NWs, it is important to develop accurate models for the oxidation kinetics in such materials.

In the popular kinetics model developed by Deal and Grove [20] for the thermal oxidation of planar silicon the intrinsic physical process of oxidation is assumed to be reaction-controlled. The chemical reaction occurs at the $SiO_2$/Si interface and hampers further oxygen diffusion through the as-formed oxide film. As the oxidation progresses, the formation of new oxide material at the interface with silicon leads to volumetric expansion, and the newly formed oxide thus displaces the older material, which consequently undergoes microstructural rearrangement.

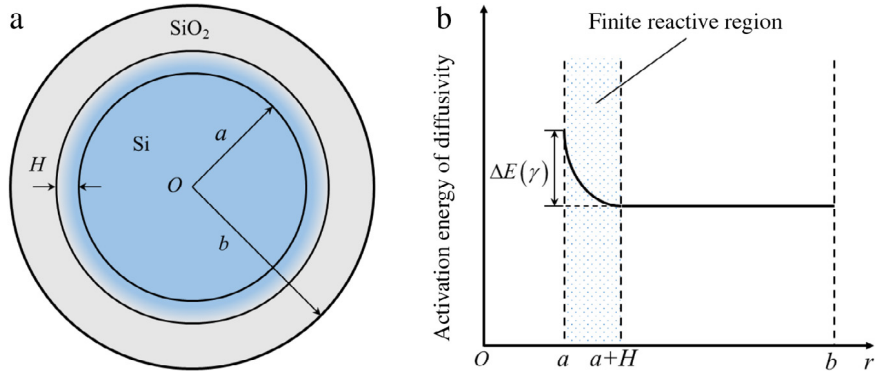

**Fig. 1.** (a) Schematic diagram of the two-dimensional cylindrical oxidation of Si NWs. (b) Profiles of the effective activation energy of $O_2$ diffusion in the oxide.

For non-planar structures, non-uniformity of oxidation induced deformation becomes significant due to the influence of surface curvature [21,22]. By considering this factor, the Deal–Grove model was later extended to the oxidation of cylindrical structures [23] and spherical particles [24]. The viscous stress associated with the non-uniform deformation of the oxide was identified as the primary factor inhibiting further oxidation and was taken into account when calculating the oxidation rate.

However, both theoretical and experimental studies have suggested that the thermal oxidation of nanowires maybe diffusion-controlled rather than reaction-controlled because the interfacial reaction is not a rate limiting step [25–27]. Specifically, first-principles calculations suggested a negligibly small activation barrier for the oxidation reaction [25], and experimental results have shown a layer-by-layer growth of oxide at the silicon surfaces, and there is also a high density at the $SiO_2/Si$ interface [26,27]. Based on these findings, Watanabe et al. [28] argued that oxygen diffusion in the oxide layer plays a leading role in governing the oxidation rate and proposed a new linear–parabolic oxidation model that does not consider rate limitation by the interfacial reaction. Cui and co-workers modified this model by considering nanoscale effects in the oxidation process and used their modified model to explain the anomalous initial oxidation process of planar Si [29] and the observed self-limiting effect in the oxidation of Si NWs [30].

Recently, atomistic simulations have suggested that oxidation does not instantaneously yield $SiO_2$. Rather a suboxide forms initially before gradually transforming to the dioxide phase [31–33]. Thus, there exists a transition layer with finite thickness between the unreacted silicon and fully reacted oxide, in which silicon exists in various oxidation states, i.e., $Si^+$, $Si^{2+}$, and $Si^{3+}$ [32]. This transition layer normally exhibits a high stress level due to stepwise chemical reactions and is expected to affect nanostructure-oxidation of other materials exhibiting suboxide formation, including tungsten nanowires (W NWs) and other transition metal nanostructures [34]. In this letter, the transition layer is referred to as the finite reactive region and is incorporated into a diffusion-controlled oxidation kinetic model to describe the oxidation behavior of nanowires. The model predictions are compared with available experimental results, showing good agreement.

To study the oxidation behavior of Si NWs, a cylindrical model is proposed, with a finite reactive region of thickness $H$ in the radial direction (see Fig. 1(a)). The inner and outer radii of the oxide shell are denoted by $a$ and $b$ with their ratio being $\gamma = b/a$. The chemical reaction process within the finite reactive region is generally too complex to be simply characterized using a physical quantity such as the surface reaction rate [35]. Here, we adopt a similar assumption to that of Watanabe et al. [28] in the cylindrical model and assume that the diffusivity in the reactive region decays exponentially due to the chemical reaction induced compressive stress. Accordingly, the effective diffusivity which is affected by the initial geometry can be expressed as a function of radial position $r$,

$$D(r, a, b) = \begin{cases} D_0, & a+H \leq r < b, \\ D_0 \exp\left[-\frac{\Delta E(\gamma)}{k_B T}\left(\frac{a+H-r}{H}\right)^2\right], \\ & a < r < a+H, \end{cases} \quad (1)$$

where $D_0$ is the diffusivity in the oxide—except for the finite reactive region, $k_B$ is the Boltzmann constant, $T$ is the oxidation temperature in K, and $\Delta E(\gamma)$ is the maximum diffusion barrier at the outer surface of the silicon core. Because high compressive stress exists in the finite reactive region during oxidation, $\Delta E(\gamma)$ is dependent upon the oxidation process in the cylindrical model, as illustrated in Fig. 1(b). Liu et al. [15] showed that the energy barrier of diffusion varies with the oxide thickness during the oxidation process. Consequently by including the results of recent molecular-dynamics simulations relating to the activation energy of diffusion [36], $\Delta E(\gamma)$ can be assumed to be

$$\Delta E(\gamma) = \Delta E_p + \Delta E_s \cdot \eta(\gamma), \quad (2)$$

where $\Delta E_p$ is the incremental diffusion barrier at the interface for planar oxidation of Si, $\Delta E_s$ is the difference in $O_2$ diffusion activation energy at the maximal density position between planar and cylindrical Si structures, and $\eta(\gamma)$ is the normalized oxide strain energy. According to Liu et al. [15], $\eta(\gamma)$ is given by

$$\eta(\gamma) = \frac{\gamma^n - 1}{(\gamma^n + 1)^2} - \frac{\gamma^{-n} - 1}{(\gamma^{-n} + 1)^2} - \frac{2n \ln(\gamma)}{(\gamma^n + 1)(\gamma^{-n} + 1)}, \quad (3)$$

where $n = 1/(1-v)$, and $v$ is Poisson's ratio with a value of 0.17 for Si NWs.

According to Fick's law [37,38] the constant flux $F_r$ of oxidant for a steady-state diffusion profile can be modeled by $F_r = -D(r) \cdot (\partial C(r)/\partial r)$, where $C(r)$ is the concentration of oxidants at the radial position $r$. Conservation of oxygen across any cylinder implies that $F_r \cdot r = $ Constant. Integrating the differential equation over $r$ from $a$ to $b$ and considering the finite reactive region ($a < r < a+H$) and the bulk region ($a+H \leq r < b$) respectively, one obtains,

$$F_a = \frac{D_0(C_b - C_i)}{a\int_a^{a+H} \frac{1}{r}\exp\left[\frac{\Delta E(\gamma)}{k_B T}\left(\frac{a+H-r}{H}\right)^2\right]dr + a\ln\left(\frac{b}{a+H}\right)}, \quad (4)$$

where $F_a$ is the oxidant flux at the outer surface of the Si core ($r = a$), and $C_b$ and $C_i$ denote the concentrations of oxidant at the outer surface of the oxide layer cylindrical shells ($r = b$) and the outer surface of Si core, respectively. According to the assumptions of Watanabe et al. [28], $C_i = 0$ can be adopted. From the Deal–Grove theory [20], the flux of the oxidant near the surface $r = b$ is taken to be $F_b = h \cdot (C^* - C_b)$, where $h$ and $C^*$ are the gas-phase transport coefficient and the equilibrium concentration in the oxide film respectively. Noting that $F_a \cdot a = F_b \cdot b = $ Constant, $F_a$ can be obtained as

$$F_a = \frac{D_0 C^*}{a \int_a^{a+H} \frac{1}{r} \exp\left[\frac{\Delta E(\gamma)}{k_B T}\left(\frac{a+H-r}{H}\right)^2\right] dr + \frac{D_0}{h}\frac{a}{b} + a \ln\left(\frac{b}{a+H}\right)}. \quad (5)$$

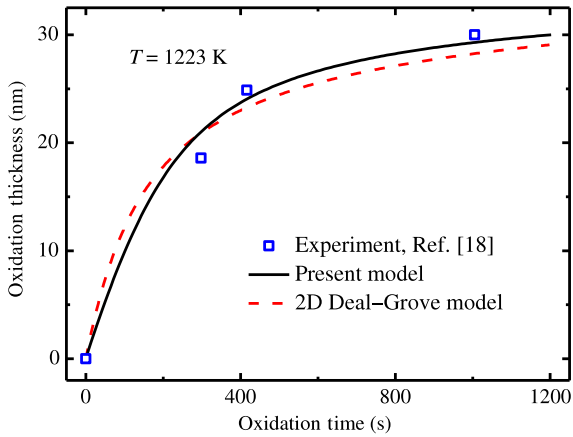

**Fig. 2.** Comparison of the model predictions and experimental results of the oxidation of a Si NW with radius $R = 25$ nm. Symbols denote experimental results, and solid and dash lines refer to the predictions by the present and 2D Deal–Grove model, respectively.

The oxidation rate at the Si/SiO$_2$ interface can be described as $dx_0/dt = F_a/N_1$ [23,30], where $x_0$ is the thickness of oxide layer and $N_1$ is the number of oxidant molecules incorporated into a unit volume of the oxide layer. Hence the kinetic rate equation for the oxidation of Si NWs which describes the change in oxide thickness over time, can be expressed as

$$\frac{dx_0}{dt} = \frac{C^*/N_1}{\frac{a}{D_0}\int_a^{a+H}\frac{1}{r}\exp\left[-\frac{\Delta E(\xi)}{k_B T}\left(\frac{a+H-r}{H}\right)^2\right]dr + \frac{1}{h}\frac{a}{b} + \frac{a}{D_0}\ln\left(\frac{b}{a+H}\right)}. \quad (6)$$

It should be noted that the present model differs from the extended 2D cylindrical Deal–Grove model by Kao et al. [23], in which the first term in the denominator of the reaction rate equation due to the interface reaction refers to the suppressed diffusion in the finite reactive region. The relationship between the oxide thickness and oxidation time can be obtained by integrating Eq. (6). Similar to the conventional Deal–Grove model, an iterative algorithm is used to numerically solve the present model. It is worth noting that when applying the 2D Deal–Grove model to small Si NWs, using this iterative method is computationally problematic. Fazzini et al. [18] have proposed an implicit algorithm to remedy this difficulty, which can avoid the sensitivity of the time step and reduce the simulation time for any initial diameter of Si NWs. However, there is no such numerical difficulty in applying the present model to nano-scaled structures. The influence of stress, if necessary, can be included into our model using stress dependent parameters similar to the conventional 2D model.

To illustrate the applicability of the developed model, several reported experimental studies on the oxidation of Si NWs are modeled. Fazzini et al. [18] measured the thermal oxidation of Si NWs with initial radius $R = 25$ nm at temperature $T = 1223$ K, the experimental data are compared with the present model predictions and are shown in Fig. 2. To simulate the experimental results, the following parameters were adopted: $H = 1$ nm, $\Delta E_p = 0.83$ eV, and $\Delta E_s = 1$ eV retrieved from Ref. [29];

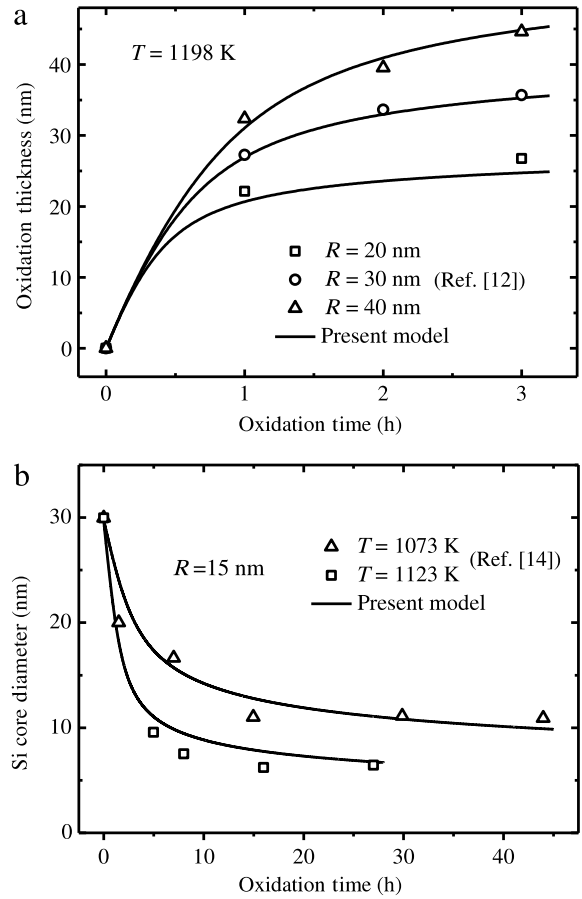

**Fig. 3.** (a) Curvature and (b) temperature dependence of the oxidation of Si NWs. Symbols denote experimental results, and solid lines show predictions of the present model.

$C^* = 0.015$ nm$^{-3}$, $D_0 = 1.92 \times 10^8$ nm$^2 \cdot$ h$^{-1}$, $h = 1.0 \times 10^{11}$ nm $\cdot$ h$^{-1}$, $N_1 = 22.5$ nm$^{-3}$, and $k_B = 1.38 \times 10^{-23}$ J $\cdot$ K$^{-1}$ determined from Ref. [18]. The prediction of the 2D Deal–Grove model of Kao et al. [23] is also included. Unlike the oxidation of planar Si, Fig. 2 shows that the oxidation of nanowires initially proceeds rapidly, before slowing down and even exhibiting self-limiting behavior for prolonged durations. It can be seen from Fig. 2 that good agreement between the model predictions and the experimental results is achieved. Although the 2D Deal–Grove model is based on the reaction dominated assumption while the present model is essentially a diffusion controlled model with a finite reactive region included, the trends shown in Fig. 2 indicate that their predictions are reasonably consistent for the studied system.

To further check the reliability of the developed model and investigate the temperature and surface curvature dependence of the self-limiting oxidation, we apply the present model to two representative sets of experimental data [12,14], as shown in Fig. 3. It can be seen that the self-limiting oxidation effect depends on the surface curvature and oxidation temperature and that the experimental data is well described by the present finite reactive region based kinetics model. For the temperature $T = 1198$ K, Fig. 3(a) shows that self-limiting behavior is predicted for the oxidation of Si NWs with different initial radii ($R = 20, 30$, and 40 nm). In addition, self-limiting oxidation is also found for Si NWs with $R = 15$ nm under two different temperatures (see Fig. 3(b), $T = 1073$ and 1123 K). Furthermore, for low temperatures and large curvatures, the finite reactive region exhibits high compressive stress and the self-limiting effect becomes stronger. In addition, it can be found that the limit oxidation thickness can be approached when the oxidation time tends to infinity for a given sample curvature and temperature condition, and the limited thickness can be predicted numerically by Eq. (6). The self-limiting oxidation effect has been used to address various technological concerns (relating to surface, size, and shape) in the production of Si NWs. The developed model can be used as the predictive tool to determine appropriate oxidation conditions and ensure the quality of obtained Si NWs and is further applicable to a broader spectrum of 2D nanostructures.

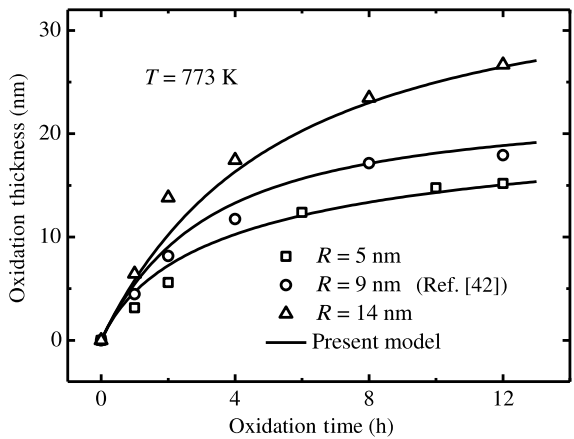

**Fig. 4.** Simulated oxide thickness of W NWs compared with experimental data. Symbols denote experimental results, and solid lines show predictions of the present model.

To further validate our present model with respect to functional nanostructures, we study the oxidation of W NWs. These materials exhibit interesting electronic, magnetic, and catalytic properties with a wide range of potential applications [39] and have been fabricated through various physic-chemical techniques including templating, thermal evaporation, and thermal oxidation [40–42]. Similar to Si NWs, self-limiting oxidation has also been observed in W NWs systems, at temperatures in the range between 673 and 773 K [42], where application of the 2D Deal–Grove model, which is based on oxide viscous flow, is problematic. Considering the similarity of the oxidation behavior in silicon and tungsten nanowires, we extend our derived 2D kinetic model to predict the self-limiting oxidation of W NWs using the expression for oxidation kinetics described in Eq. (6).

In order to represent the experimental conditions, model parameters (i.e., $C^* = 7.5 \times 10^{-5}$ nm$^{-3}$, $D_0 = 3.6 \times 10^{10}$ nm$^2 \cdot$h$^{-1}$, $h = 1.0 \times 10^{11}$ nm $\cdot$ h$^{-1}$, and $N_1 = 16.5$ nm$^{-3}$) were retrieved from Ref. [42] and the Poisson's ratio was taken from Ref. [43]. The temperature $T$ was taken as 773 K, as used in the experimental method. Parameters $\Delta E_p = 0.89$ eV and $\Delta E_s = 0.14$ eV were determined on the basis of reported experimental data. The oxide thickness for conditions of different initial diameters as predicted by the present model is plotted as the solid lines in Fig. 4. For comparison, the experimental results by You and Thong [42] are included. It is evident from Fig. 4 that the current model predictions are consistent with the experimental oxidation data, and can reproduce the observed self-limiting behavior. As with Si NWs, the self-limiting effect is stronger for larger curvature values.

In this letter, we extend Watanabe's diffusion-controlled one-dimensional kinetics model of planar Si oxidation to 2D cylindrical situation by introducing the concept of a finite reactive region and further considering the effects of reaction induced compressive stress on the oxidation rate in the finite region. The developed model can be easily applied to nanowires of smaller dimensions without numerical difficulty. Good agreement between the predictions given by the present model and the 2D Deal–Grove model and the experiential results of Si and W NWs implies that the predictions of reaction and diffusion based models may be similar for certain systems. In addition, the experimentally observed curvature and temperature dependent oxidation behavior of nanowires is also predicted. The developed model allows us to quantify nanowires oxidation kinetics and may facilitate the optimization of 2D nanostructure fabrication via thermal oxidation processes.

### Acknowledgments

The authors are grateful for the financial support of this work by the National Natural Science Foundation of China (11472149), and the Tsinghua University Initiative Scientific Research Program (2014z22074). Discussion of the numerical method for the implicit model with Pier-Francesco Fazzini is greatly appreciated.